\documentclass[twocolumn,preprintnumbers,superscriptaddress,nofootinbib,aps,prl,floatfix]{revtex4-2}
\pdfoutput=1
\usepackage[utf8]{inputenc}
\usepackage{amsmath,amssymb}
\usepackage{epsfig}
\usepackage{comment}
\usepackage{bbm}
\usepackage{graphicx}
\usepackage[usenames,dvipsnames]{color}
\usepackage{float}
\usepackage{bbold}
\usepackage[colorlinks,citecolor=blue]{hyperref}
\usepackage[normalem]{ulem}

\usepackage{color}
\usepackage{subfigure, orcidlink}

\begin{document}
\title{Imprint of matter-antimatter asymmetry on collapsing domain walls}

\author{Dipendu Bhandari \orcidlink{https://orcid.org/0000-0003-1321-6006}}
\email{dbhandari@iitg.ac.in}
\affiliation{Department of Physics, Indian Institute of Technology Guwahati, Assam 781039, India}

\author{Debasish Borah \orcidlink{https://orcid.org/0000-0001-8375-282X}}
\email{dborah@iitg.ac.in}
\affiliation{Department of Physics, Indian Institute of Technology Guwahati, Assam 781039, India}

\author{Indrajit Saha \orcidlink{https://orcid.org/0000-0002-7459-0838}}
\email{s.indrajit@iitg.ac.in}
\affiliation{Department of Physics, Indian Institute of Technology Guwahati, Assam 781039, India}

\begin{abstract}
Spontaneous breaking of discrete symmetries play non-trivial role in many well-motivated particle physics models. However, it leads to a network of cosmologically unwanted domain walls (DWs) which can be made unstable by introducing a bias term in the scalar potential.
In this letter, we provide a novel origin of such bias terms at finite temperature due to radiative corrections from a Dirac fermion with large asymmetry $\sim \mathcal{O}(0.1)$ in its number density. In addition to getting a new viable region of parameter space for collapsing DWs not explored previously and resulting gravitational waves (GWs) accessible at future experiments, the viability of the scenario crucially depends on the temperature of asymmetry generation too. This provides a unique way of probing both the amount of asymmetry and the corresponding temperature via future observations of GWs from collapsing DWs. The large asymmetry in the Dirac fermion can also have interesting implications for the observed baryon asymmetry as well as dark matter and large neutrino asymmetry. \\~\\

\end{abstract}

\maketitle

\noindent
\textit{{Introduction:}} Formation of topological defects like domain walls (DWs) is ubiquitous in several particle physics models with spontaneous discrete symmetry breaking. The simplest among them is the one which arises after the spontaneous breaking of a $\mathcal{Z}_2$ symmetry by the vacuum expectation value (VEV) of a scalar field. Such DWs, if stable, can quickly dominate the energy density of the Universe leading to a cosmological catastrophe. Such a catastrophe can be avoided if the DWs are made unstable by introducing a small explicit symmetry breaking through a bias term in the scalar potential \cite{Zeldovich:1974uw, Vilenkin:1981zs, Sikivie:1982qv, Gelmini:1988sf, Larsson:1996sp}. Such a bias term creates a pressure difference across the walls leading to their collapse and release of energy in the form of stochastic gravitational waves (GWs). While it is possible to add such explicit $\mathcal{Z}_2$ symmetry breaking terms in an \textit{ad-hoc} manner, their origin can be related to higher-dimensional operators suppressed by the scale of quantum gravity \cite{Rai:1992xw, Lew:1993yt}, radiative corrections from dark fermions or heavy neutral leptons taking part in seesaw \cite{Zhang:2023nrs, Zeng:2025zjp, Borah:2025bfa, Borah:2026kfo}, or a lepton parity~\cite{Ma:2025bjf}.

In this letter, we propose a novel scenario where the bias term required for the collapse of DWs is generated primarily from radiative corrections induced by a highly degenerate fermion species. This allows a direct link between the bias term and the chemical potential or particle-antiparticle asymmetry of the fermion species. While the asymmetry-dependent contribution to the bias arises from finite-temperature effects, there exists a zero-temperature contribution to the bias at radiative level which is independent of the asymmetry or the chemical potential. We find the parameter space of the model where the asymmetry-dependent contribution dominates over the latter. This requires $\mathcal{O}(0.1)$ asymmetry in the fermion species such that the GWs from collapsing DWs remain within future experimental reach. Interestingly, the collapse of the DWs is not only sensitive to the asymmetry, but also the epoch when such asymmetry is generated or injected into the bath. This provides a unique way of probing both the asymmetry and the scale of its production via their effects on collapsing DWs and resulting GWs. Although the required $\mathcal{O}(0.1)$ asymmetry is much larger compared to the observed matter-antimatter asymmetry in visible or baryonic matter $\sim \mathcal{O}(10^{-10})$ \cite{Planck:2018vyg}, it can still have interesting implications for light neutrinos or dark matter (DM) which can store such large asymmetry with observable consequences associated with them. \\

\noindent
\textit{{Biased domain walls with degenerate fermions:}} In order to demonstrate the idea, we consider a $\mathcal{Z}_2$-odd scalar field $\phi$ with a scalar potential 
\begin{equation}
    V (\phi) = -\frac{\mu^2_\phi}{2} \phi^2 + \frac{\lambda_\phi}{4} \phi^4.
\end{equation}
For $\mu^2 >0$, the scalar field $\phi$ acquires a non-zero VEV $v_\phi\simeq\sqrt{\mu^2_\phi/\lambda_\phi}$, spontaneously breaking the $\mathcal{Z}_2$ symmetry. This leads to the formation of domain walls in the early Universe \cite{Zeldovich:1974uw, Kibble:1976sj, Vilenkin:1981zs,Saikawa:2017hiv,Roshan:2024qnv} which, if stable, would dominate the energy density of the Universe, creating conflict with standard cosmological observations related to the cosmic microwave background (CMB) and the big bang nucleosynthesis (BBN). Assuming the walls to be formed after inflation, the simplest way to make them disappear is to introduce a small pressure difference \cite{Zeldovich:1974uw, Vilenkin:1981zs, Sikivie:1982qv, Gelmini:1988sf, Larsson:1996sp}, often referred to as the bias term, which breaks $\mathcal{Z}_2$ explicitly. In most parts of the relevant literature, such bias terms are typically written as odd powers of the $\mathcal{Z}_2$-odd scalar with small coefficients in the scalar potential, which can arise either from higher dimensional operators or radiative corrections. The role of such bias terms is to lift the degeneracy between the two minima, ensuring the eventual collapse of the domain wall network before they dominate or overclose the Universe. In this work, we propose the origin of such bias term from radiative corrections to the scalar potential due to a degenerate fermion $\chi$ which couples to the scalar field $\phi$ via Yukawa interactions $-y \bar{\chi} \chi \phi$. Given that the vector-like fermion $\chi$ also has a bare mass given by $m_0$, its field-dependent mass is $m_\chi (\phi) = m_0 + y \phi$.

The explicit $\mathcal{Z}_2$-breaking Yukawa interactions can lead to bias terms in the scalar potential in two different ways namely, via zero-temperature radiative corrections and finite-temperature corrections. The zero-temperature correction at one-loop level, is given by the Coleman-Weinberg correction \cite{Coleman:1973jx}. Assuming $\langle \phi \rangle=-v_\phi$ to be the true minima such that the one-loop Coleman-Weinberg bias term at $T=0$ can be written as 
\begin{align}
    \Delta V_{\rm cw}= V_{\rm cw}(+v_\phi) - V_{\rm cw} (-v_\phi)
\end{align}
where 
\begin{align}
    V_{\rm cw}( \pm v_\phi)&  = \sum_i \frac{-1}{64\pi^2}n_i(m_{\chi}(\pm v_\phi))^4 \nonumber \\  
   & \times \Big\{{\rm Log}\Big(\frac{(m_{\chi}( \pm v_\phi))^2}{\mu^2_r}\Big) -3/2\Big\}
\end{align}
where $n_i=4$ is the degrees of freedom of $\chi$ and $\mu_r$ is the renormalisation scale, taken to be $v_\phi$. On the other hand, the finite-temperature correction can be written as 
\begin{equation}
    \Delta V_T  = V_T(+v_\phi)-V_T(-v_\phi) = \frac{2 m_0 y v_\phi T^2}{\pi^2}\mathcal{G}(\mu/T)
\end{equation}
where $\mathcal{G}(\mu/T)$ is a function dependent on temperature as well as the chemical potential $\mu$ of $\chi$ the details of which can be found in \textit{Appendix}. It should be noted that we use the full finite-temperature potential in our numerical analysis without any approximation.\\

\noindent
\textit{{Gravitational waves from collapsing domain walls:}} Once the discrete $\mathcal{Z}_2$ symmetry is spontaneously broken, a network of domain walls form. Such DWs are characterized by their surface energy density, commonly referred to as the domain wall tension  $\sigma_{\rm DW}=\frac{4}{3}\sqrt{\frac{\lambda_\phi}{2}}v_\phi^3\simeq\frac{2}{3}m_{\phi}v_{\phi}^2$ where $m_{\phi}=\sqrt{2\lambda_\phi}v_\phi$. Due to the presence of the bias terms discussed above, the domain walls become unstable and eventually collapse, while emitting a stochastic gravitational wave background \cite{Vilenkin:1981zs, Gelmini:1988sf, Larsson:1996sp, Hiramatsu:2013qaa, Hiramatsu:2012sc, Kadota:2015dza, Saikawa:2017hiv, Chen:2020wvu, Roshan:2024qnv,Bhattacharya:2023kws,Paul:2024iie,Ma:2025bjf}. The annihilation temperature $T_{\rm ann}$ of DWs can be estimated by equating the bias term or pressure difference with the domain wall's tension force. For the finite-temperature part dominating the bias term, we get 
\begin{align}
   & 1.66 \sqrt{g_*(T)}\frac{1}{M_{\rm pl}}= C_1 \frac{  \mathcal{G}(\mu/T)}{0.94\pi^2 \sqrt{\lambda_\phi}C_{\rm ann}\mathcal{A}}
\end{align}
from the equality of pressure difference and the tension force. Here, $C_{\rm ann}\simeq 2$ is a dimensionless constant, $g_*$ is the relativistic degrees of freedom, $\mathcal{A}\simeq 0.8$ is the area parameter \cite{Hiramatsu:2013qaa} and $C_1=\frac{m_0 y}{v_\phi^2}$. Using the relation between chemical potential $\mu$ and comoving density $Y_{\Delta \chi}$, we get $C_1=10^{-22}$ by using $Y_{\Delta\chi}^{\rm min}$=0.16. Here, $Y_{\Delta\chi}^{\rm min}$ is the minimum comoving asymmetry in $\chi$ required for DWs to annihilate at a temperature when the asymmetry is generated or injected, denoted by $T_{\rm inj}$. 

\begin{figure}
    \centering
    \includegraphics[width=\linewidth]{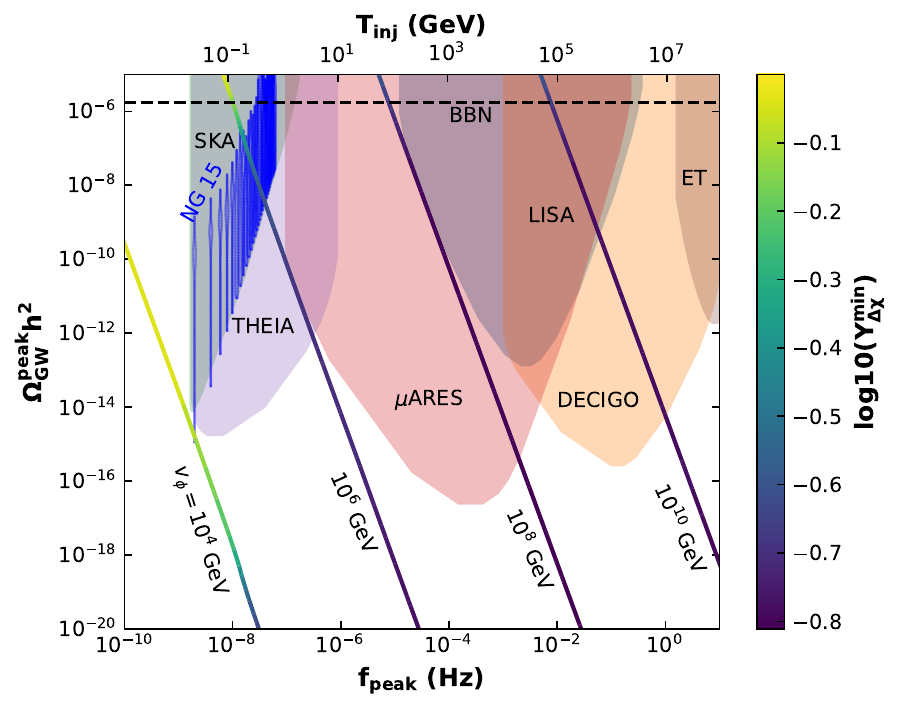}
    \caption{Gravitational wave peak amplitude as a function of peak frequency. The colored curves correspond to symmetry breaking occurring at different values of $v_{\phi}$. The color scale represents the minimum amount of asymmetry required at the injection temperature, indicated on the top axis. Here, $C_1=10^{-22}$ is fixed.}
    \label{fig:1}
\end{figure}
\begin{figure}
    \centering
    \includegraphics[width=\linewidth]{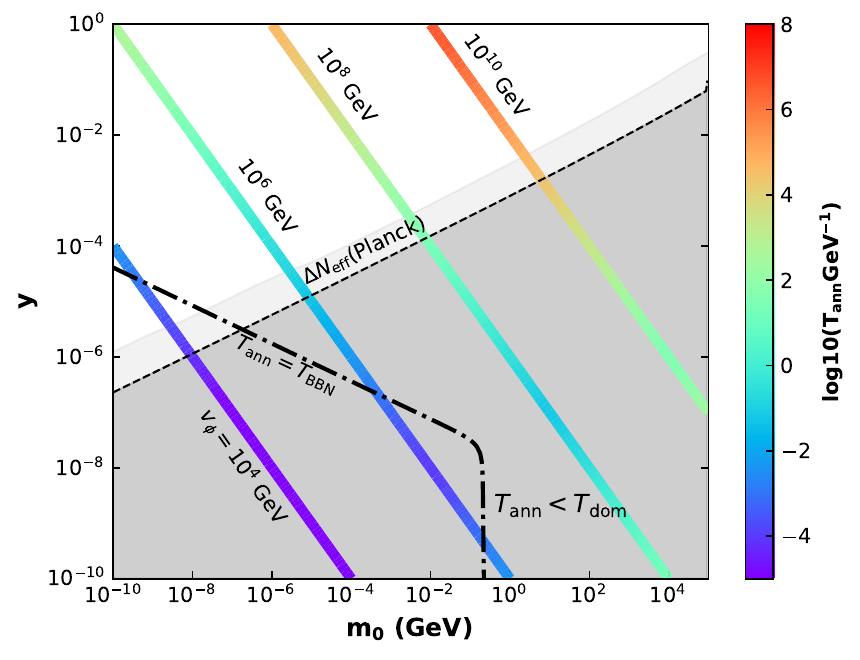}
    \caption{Parameter space in the $y$--$m_0$ plane corresponding to Fig.~\ref{fig:1}. Keeping $C_1$ fixed, the different colored lines represent different values of $v_\phi$. The color scale indicates the annihilation temperature arising from the zero-temperature bias. This determines the minimum temperature before which the asymmetry must be injected to trigger an earlier annihilation.}
    \label{fig:2}
\end{figure}

The bias term in the scalar potential also has other phenomenological constraints. A very small bias term will allow domain walls to dominate the energy density of the Universe at a temperature given by \cite{Saikawa:2017hiv}
\begin{align}
    T_{\rm dom} &= 2.86\times10^{-6} \left (\frac{g_*}{10} \right)^{-1/4} \left (\frac{\mathcal{A}}{0.8} \right )^{1/2} \nonumber \\
    & \times \left (\frac{\sigma_{\rm DW}}{10^9\, {\rm GeV^3}} \right )^{1/2} {\rm GeV}.
\end{align}
For domain wall annihilation to occur before their domination $T_{\rm ann} > T_{\rm dom}$, we have a lower bound on the bias term. Another lower bound on the bias term arises by demanding the domain walls to annihilate before the BBN epoch $T_{\rm ann} > T_{\rm BBN}$ such that the light nuclei abundances do not get altered. Finally, the bias term can not be arbitrarily large either as it would otherwise prevent the percolation of both the vacua separated by DWs. This leads to an upper bound on the bias term $V_{\rm bias} < 0.795 V_0$ \cite{Saikawa:2017hiv} with $V_0$ being the height of the potential barrier separating the two minima. \\

Fig. \ref{fig:1} shows the peak amplitudes and peak frequencies of the GW spectra for different values of the symmetry breaking scale of $v_\phi$ and a fixed value of $C_1$. Here, we also fixed the value of $\lambda_\phi$ to be $0.1$ and for the rest of the analysis. The collapse of the DWs are assumed to occur dominantly due to the finite-temperature bias term dependent on the asymmetry of $\chi$, indicated by the color code in Fig. \ref{fig:1}. Clearly, the peak amplitude of GW decrease with decreasing comoving asymmetry, as expected. As mentioned above, the annihilation of DWs and hence the resulting GW spectrum are also sensitive to the temperature at which the asymmetry is injected $T_{\rm inj}$, shown in the top horizontal axis of Fig. \ref{fig:1}. Higher injection temperature corresponds to early annihilation and hence higher GW peak frequencies, as expected. The dashed horizontal line corresponds to the upper limit on new contributions to the effective number of relativistic degrees of freedom, $\Delta{N}_{\rm eff}$ or dark radiation from gravitons or GW. A sufficiently large GW energy density behaves as dark radiation and contributes to the total radiation content of the Universe, thereby modifying the expansion rate during BBN and recombination. This leads to an upper bound on the GW energy density, which can be expressed as \cite{Caprini:2018mtu}
\begin{eqnarray}
    \Omega_{\rm GW}h^2\lesssim 5.6\times10^{-6}\Delta{N}_{\rm eff}.
\end{eqnarray}
Shaded regions in Fig. \ref{fig:1} correspond to the sensitivities of various future GW experiments like LISA~\cite{2017arXiv170200786A}, ET~\cite{Punturo_2010}, THEIA~\cite{Garcia-Bellido:2021zgu}, $\mu$ARES~\cite{Sesana:2019vho}, SKA~\cite{Weltman:2018zrl}, DECIGO~\cite{Adelberger:2005bt} while the blue violin-shaped points show the NANOGrav (NG) results \cite{NANOGrav:2023gor}.

\begin{figure*}
    \centering
    \includegraphics[scale=0.40]{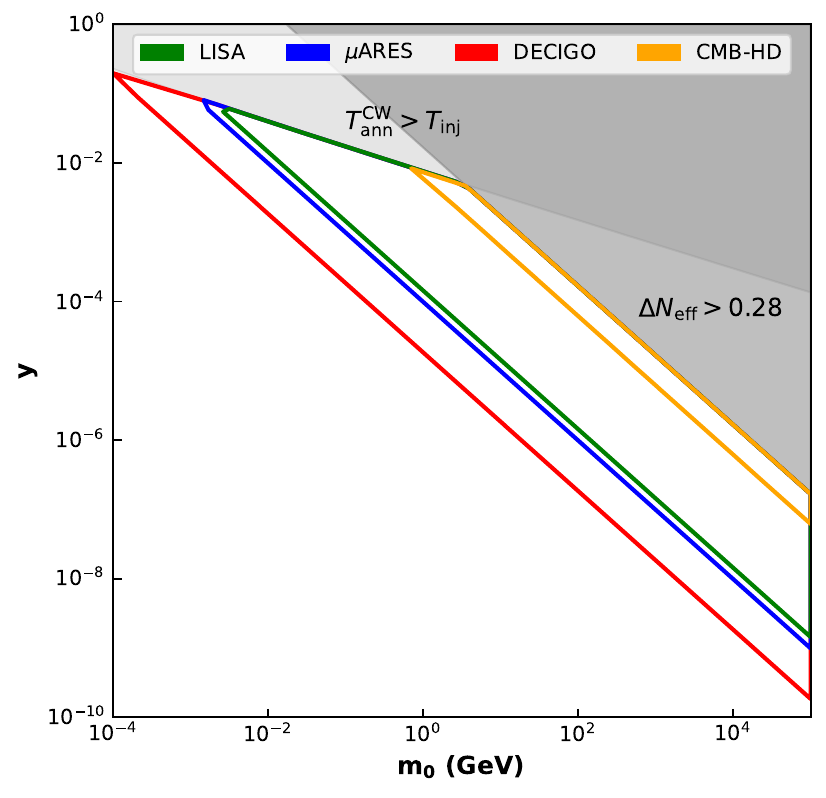}
    \includegraphics[scale=0.40]{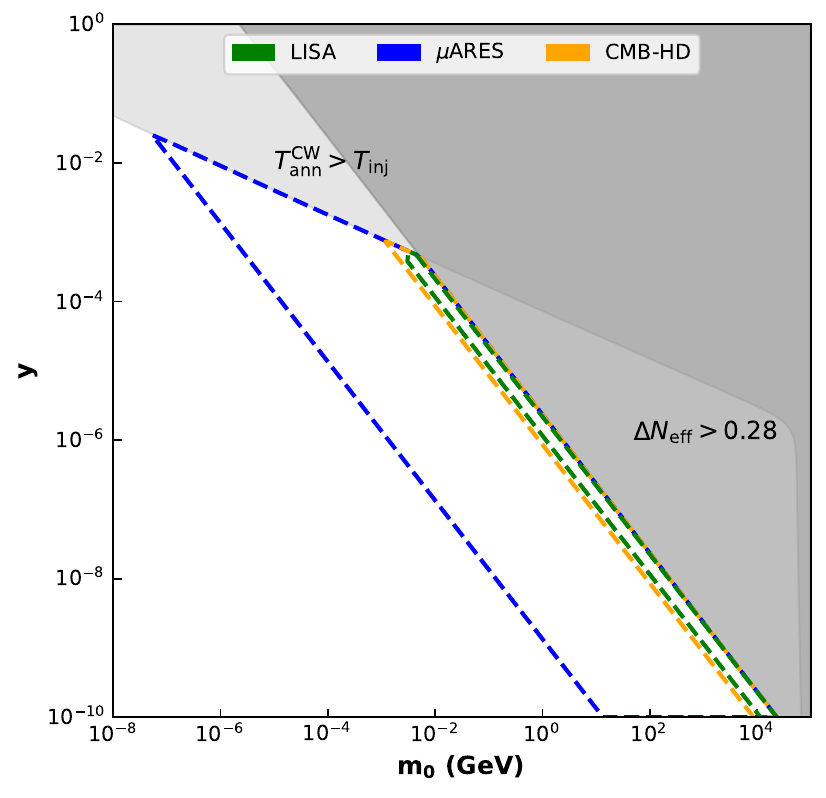}
    \includegraphics[scale=0.40]{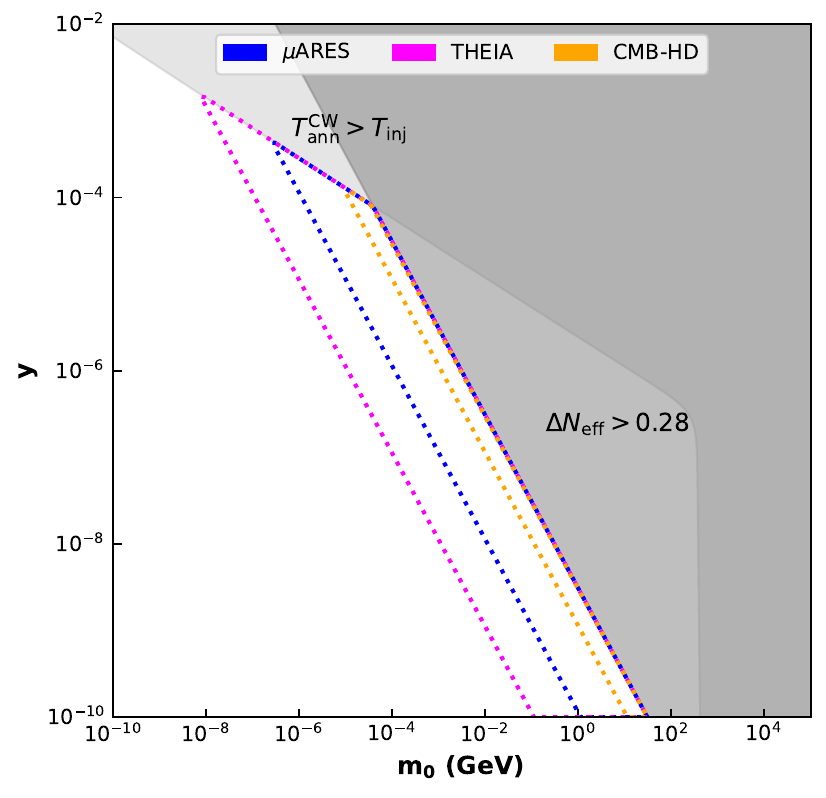}
    \caption{Parameter space in the $y$--$m_0$ plane. The contours correspond to fixed asymmetry injection temperatures of $10^5$~GeV (solid), $130$~GeV (dashed), $1$~GeV (dotted). The regions within the contours can be probed by gravitational wave experiments as well as $\Delta N_{\rm eff}$ experiments, indicated by different colors. The comoving asymmetry $Y_{\Delta \chi}$ is kept at the maximum allowed value $\sim \mathcal{O}(0.1)$ for each of the injection temperatures.}
    \label{fig:3}
\end{figure*}


Fig. \ref{fig:2} shows the parameter space in $y-m_0$ plane for different values of the symmetry breaking scale $v_\phi$. The color code in different bands for fixed $v_\phi$ corresponds to the annihilation temperature of DWs considering only the zero-temperature bias term generated from the Coleman-Weinberg corrections. The dark grey-shaded region corresponds to $T_{\rm ann} < T_{\rm dom}$ and hence disfavored. The domination temperatures are $6.42\times10^{-5}$, $4.52\times10^{-2}$, $2.84\times10^{1}$, and $2.73\times10^{4}\,\mathrm{GeV}$, corresponding to increasing values of $v_\phi$ shown in Fig. \ref{fig:2}. So, the edges of the dark grey region corresponds to the temperature above which the asymmetry in $\chi$ has to be injected such that DWs can annihilate before dominating. Thus, the parameter space disfavored with zero-temperature bias can become viable with finite-temperature bias provided the required $\chi$ asymmetry is generated at a temperature above the minimum temperature corresponding to the edges of the dark grey region of Fig. \ref{fig:2}. The light grey-shaded region is disfavored from PLANCK 2018 bound $N_{\rm eff}=2.99^{+0.34}_{-0.33}$ at $2\sigma$ CL \cite{Planck:2018vyg} including baryon acoustic oscillation (BAO) data. This leads to a stronger constraint of injecting the asymmetry at higher temperatures ($1.72\times10^{-4}$, $0.12$, $6.94\times10^1$ and $6.81\times10^4$ GeV) so that $\Omega_{\rm GW} h^2$ is sufficiently suppressed to satisfy this bound. The parameter space in the lower left corner enclosed by the dot-dashed line corresponds to $T_{\rm ann} < T_{\rm BBN}$ and hence disfavored.

Fig. \ref{fig:3} shows the summary of our results in $y-m_0$ parameter space highlighting the newly allowed detectable region introduced by finite-temperature effects including the chemical potential or asymmetry of Dirac fermion $\chi$ with $\mathcal{Z}_2$-breaking Yukawa coupling to $\phi$. The left, middle and right panel plots of Fig. \ref{fig:3} correspond to three different injection temperatures namely $T_{\rm inj}=10^5$ GeV, $130$ GeV and $1$ GeV respectively. The light grey colored regions in all of these figures correspond to the parameter space where the annihilation of DWs is governed primarily by the zero-temperature bias terms. The dark grey shaded regions are ruled out from PLANCK 2018 bound on $N_{\rm eff}$ mentioned earlier. The green, blue, red and pink colored contours enclose the parameter space which can be probed in future experiments like LISA, $\mu$ARES, DECIGO and THEIA respectively. The orange colored contour encloses the parameter space which can be probed at CMB-HD~\cite{CMB-HD:2022bsz} experiment sensitive to enhanced dark radiation or $N_{\rm eff}$.
\\

\noindent
\textit{{Implications for the observed matter-antimatter asymmetry and dark matter:}} As discussed above, the required comoving asymmetry $Y_{\Delta \chi}$ is very large $\sim \mathcal{O}(0.1)$ in order to play the leading role in the collapse of DWs. While this much larger compared to the observed baryon asymmetry in the Universe $\sim \mathcal{O}(10^{-10})$, it is also much larger for asymmetric dark matter $\chi$ with mass $m_{\chi}=m_0+y v_\phi$ in the range discussed above. Nevertheless, we can have several interesting possibilities where such large asymmetry stored in singlet Dirac fermion $\chi$ can have other observable consequences in addition to GWs from collapsing DWs.

Firstly, such vector-like singlet fermions appear in seesaw models \cite{Barman:2022yos, Barman:2023fad, King:2023cgv, Borah:2026kfo} for light Dirac neutrinos, opening up their decays into the light neutrinos while also transferring the asymmetry. If such decays occur below the sphaleron decoupling epoch $T_{\rm sph} \sim 131.7$ GeV, the overproduction of baryon asymmetry can be avoided, while storing the large asymmetry in neutrinos. Such large neutrino asymmetry can lead to an enhanced $N_{\rm eff}$ \cite{Escudero:2022okz, Borah:2024xoa}. Additionally, such large neutrino asymmetry also has the potential to alleviate the recently reported Helium anomaly \cite{Matsumoto:2022tlr, Burns:2022hkq}.

Secondly, decay of Dirac fermion $\chi$ into leptons and dark sector can occur together providing a viable cogenesis mechanism \cite{Borah:2026xxx, Borah:2025dka, Bandyopadhyay:2025hoc}. The relative branching ratio as well as partial decay widths can be tuned in such a way that the lepton sector asymmetry generated by sphaleron decoupling epoch gives rise to the desired baryon asymmetry \cite{Borah:2022uos} while subsequent decays into leptons and dark matter can lead to large neutrino asymmetry and asymmetric dark matter respectively. \\

\noindent
\textit{{Conclusion:}} We have proposed a novel solution to the the problem of cosmologically unwanted stable domain walls which arise in several particle physics models with spontaneous discrete symmetry breaking. Considering the simplest example of a $\mathcal{Z}_2$-odd scalar $\phi$, we introduce an explicit symmetry-breaking term involving $\phi$ and a singlet Dirac fermion $\chi$ with a large in-built asymmetry in its number density. Considering the radiative corrections to the scalar potential at zero and finite temperatures, we identify the parameter space of the model where finite-temperature corrections dependent on $\chi$ asymmetry play the dominant role in the collapse of $\mathcal{Z}_2$ domain walls. In this parameter space, zero-temperature corrections or finite-temperature corrections independent of the chemical potential alone can not provide a cosmologically viable solution to the domain wall problem. While a large $\chi$ asymmetry $\sim \mathcal{O}(0.1)$ is required to have a visible effect on the collapse of DWs, we also find interesting dependence of this collapse on the scale of asymmetry generation. In the parameter space where asymmetry-dependent finite-temperature bias term dominates, the DWs annihilate at a temperature when the asymmetry is generated or injected into the bath. Since annihilation temperature and the strength of the bias term play crucial role in the resulting gravitational wave peak amplitude and peak frequency, our proposed scenario provides a unique way of probing not only the amount of matter-antimatter asymmetry but also the corresponding injection temperature. We show that a large part of the parameter space of the model with varying asymmetry injection temperatures can be probed with future observations of stochastic GWs at experiments like LISA, DECIGO as well as future CMB experiments sensitive to the enhanced dark radiation. Our proposed framework also opens up several future directions in studying the role of such large asymmetry stored in $\chi$ in generating the observed baryon asymmetry as well as asymmetries in neutrino and dark matter sectors with interesting observational aspects. While we have not studied the origin of $\chi$ asymmetry here, one minimal way of generating it is the Affleck-Dine framework \cite{Affleck:1984fy}. We leave such UV completions and other phenomenological relevance of $\chi$ asymmetry to future studies.
\\

\noindent
\textit{{Acknowledgments:}} 
The work of D.B. is supported by the Science and Engineering Research Board (SERB), Government of India grant CRG/2022/000603. The work of I.S. is supported by SERB, Government of India grant CRG/2022/000603.

\newpage
\begin{center}
    {\bf  \large Appendix}
\end{center}
\appendix

\section{Finite-temperature bias with non-zero chemical potential}
The thermal potential correction due to a vector-like fermion $\chi$ can be written as
\begin{align}
V_T^\chi &= -4\frac{1}{\beta} \int \frac{d^3 k}{(2\pi)^3} 
\ln \left( 1 + e^{-\beta \omega_k} \right),
\quad 
\omega_k = \sqrt{k^2 + m_\chi^2} \nonumber 
\end{align}
where $\beta = 1/T$, and the factor of $4$ corresponds to the degrees of freedom of $\chi$ in absence of chemical potential.
In the presence of a finite chemical potential $\mu$, the thermal corrections must be computed separately for $\chi$ and $\bar{\chi}$. The contributions to the scalar potential from $\chi$ and $\bar{\chi}$ are given by
\begin{align}
V_T^{\chi/\bar{\chi}} &= -2\frac{1}{\beta} \int \frac{d^3 k}{(2\pi)^3} 
\ln \left( 1 + e^{-\beta (\omega_k \mp \mu)} \right).
\end{align}
The total thermal correction can be written as
\begin{align}
V_T 
= & -\frac{2}{\beta} \int \frac{d^3 k}{(2\pi)^3} 
\left[
\ln \big( 1 + e^{-\beta (\omega_k - \mu)} \right)\nonumber\\
& + \ln \left( 1 + e^{-\beta (\omega_k + \mu)} \right)
\Big].
\end{align}
Now, we expand the logarithmic part
\begin{align}
 V_T 
& = -\frac{2}{\beta} \sum_{n=1}^{\infty} \frac{(-1)^{n+1}}{n}
\, 2\cosh(n\beta\mu)
\int \frac{d^3 k}{(2\pi)^3}
e^{-n\beta \omega_k}. \nonumber 
\end{align}
We can write the integration part as
\begin{align}
    \int \frac{d^3 k}{(2\pi)^3}
e^{-n\beta \omega_k} = & \int \frac{d^3 k}{(2\pi)^3}
e^{-n\beta \sqrt{k^2 + m_\chi^2}}\nonumber\\
= & \frac{m^2_\chi}{2\pi^2\, n\beta}
K_2(n\beta m_\chi) 
\end{align}
where, $K_2(n\beta m_\chi)
= \int_0^\infty dx \,
e^{-n\beta m_\chi \cosh x}
\cosh(2x)$. Then, we can express the correction as
\begin{align}
V_T =& -\frac{2}{\pi^2} \frac{m^2_\chi}{\beta^2}
\sum_{n=1}^{\infty}
(-1)^{n+1}
\frac{\cosh(n\beta\mu)}{n^2}
K_2(n\beta m_\chi) \label{full_VT} \\
 =& -\frac{2 m^2_\chi}{\pi^2\beta^2}
 \frac{1}{2}
\int_0^\infty dx \,
\cosh(2x)
\sum_{n=1}^{\infty}
\frac{(-1)^{n+1}}{n^2} \nonumber \\
& \times \Big[
e^{-n(\beta m_\chi \cosh x - \beta\mu)}+
e^{-n(\beta m_\chi \cosh x + \beta\mu)}
\Big].
\end{align}
We can write
\begin{align}
\sum_{n=1}^{\infty}
(-1)^{n+1}
\frac{
\left[
e^{-(\beta m_\chi \cosh x - \beta\mu)}
\right]^n
}{n^2}
&=
- \sum_{n=1}^{\infty}
\frac{
\left[
- e^{-(\beta m_\chi \cosh x - \beta\mu)}
\right]^n
}{n^2}\nonumber
\\[6pt]
&=
- \operatorname{Li}_2
\!\left(
- e^{-(\beta m_\chi \cosh x - \beta\mu)}
\right) \nonumber 
\end{align}
where $\operatorname{Li}$ is the PolyLog function. Finally, we can write the correction to the scalar potential as
\begin{align}
V_T 
&= \left(\frac{m^2_\chi}{\pi^2\beta^2}\right)
\int_0^\infty dx \, \cosh(2x)
\Bigg\{
\operatorname{Li}_2
\!\left[-e^{-(\beta m_\chi \cosh x - \beta\mu)}\right]
\nonumber\\
&\hspace{4cm}
+
\operatorname{Li}_2
\!\left[-e^{-(\beta m_\chi \cosh x + \beta\mu)}\right]
\Bigg\}. \nonumber 
\end{align}
For high temperature approximation, we can expand the Bessel function as
\begin{align}
K_2(n\beta m_\chi)
& =
\frac{2}{n^2 \beta^2 m^2_\chi}
-\frac{1}{2}
+\frac{n^2 \beta^2 m^2_\chi}{16}
\left(
2\gamma_E -1 + 2\ln\frac{n\beta m_\chi}{2}
\right) \nonumber\\
& + \mathcal{O}((n\beta m_\chi)^4)
\end{align}
Keeping the first two leading terms, we can write the Eq.~\eqref{full_VT}
\begin{align}
V_T
= &
-\frac{2m^2_\chi}{\pi^2\beta^2} 
\sum_{n=1}^{\infty}
(-1)^{n+1}
\frac{\cosh(n\beta\mu)}{n^2}
\left(
\frac{2}{n^2 \beta^2 m^2_\chi}
-\frac{1}{2}
\right) \nonumber\\
= &
-\frac{2m^2_\chi}{\pi^2\beta^2}
\Bigg[
\sum_{n=1}^{\infty}
(-1)^{n+1}
\frac{\cosh(n\beta\mu)}{n^4}
\frac{2}{\beta^2 m^2_\chi}\nonumber\\
&-
\sum_{n=1}^{\infty}
(-1)^{n+1}
\frac{\cosh(n\beta\mu)}{n^2}
\frac{1}{2}
\Bigg].
\end{align}
Using property of PolyLog function ($\operatorname{Li}$),
$\sum_{n=1}^{\infty}
\frac{(-1)^{n+1} e^{nx}}{n^t}
=- \operatorname{Li}_t(-e^{x})$, we can write
\begin{align}
\sum_{n=1}^{\infty}
(-1)^{n+1}
\frac{\cosh(nx)}{n^t}
&=
\frac{1}{2}
\sum_{n=1}^{\infty}
\frac{(-1)^{n+1}}{n^t}
\left(e^{nx}+e^{-nx}\right)
\nonumber\\
&=
-\frac{1}{2}
\left[
\operatorname{Li}_t(-e^{x})
+
\operatorname{Li}_t(-e^{-x})
\right]. \nonumber 
\end{align}

Now, thermal correction can be expressed as
\begin{align}
V_T
&=
-\frac{2m^2_\chi}{\pi^2\beta^2}
\Bigg[
\frac{2}{\beta^2 m^2_\chi}
\left(
-\frac{1}{2}
\left[
\operatorname{Li}_4(-e^{\beta\mu})
+
\operatorname{Li}_4(-e^{-\beta\mu})
\right]
\right)
\nonumber\\
&\qquad
-
\frac{1}{2}
\left(
-\frac{1}{2}
\left[
\operatorname{Li}_2(-e^{\beta\mu})
+
\operatorname{Li}_2(-e^{-\beta\mu})
\right]
\right)
\Bigg].
\end{align}
Denoting $\chi$ mass, $m_\chi(\phi)=m_0 +y \phi$, at $\langle \phi \rangle=+v_\phi$ as $m_+=m_0 +y v_\phi$ and at $\langle \phi \rangle= -v_\phi$ as $m_-=m_0 -y v_\phi$, the thermal bias term can be written as 
\begin{align}
    \Delta V_T & = V_T(+v_\phi)-V_T(-v_\phi) \nonumber \\
                & = - \frac{1}{2\pi^2\beta^2}\left[
\operatorname{Li}_2(-e^{\beta\mu})
+
\operatorname{Li}_2(-e^{-\beta\mu})
\right] (m_+^2-m_-^2) \nonumber \\
 & =  \frac{2 m_0 y v_\phi T^2}{\pi^2}\left[-
\operatorname{Li}_2(-e^{\mu/T})
-
\operatorname{Li}_2(-e^{-\mu/T})
\right] \nonumber \\
& =  \frac{2 m_0 y v_\phi T^2}{\pi^2}\mathcal{G}(\mu/T)
\end{align}
where, $\mathcal{G}(\mu/T)=\left[-
\operatorname{Li}_2(-e^{\mu/T})
-
\operatorname{Li}_2(-e^{-\mu/T})
\right]$. Writing the comoving asymmetry in $\chi$ as
\begin{align}
    Y_{\Delta\chi}  =\frac{n_{\Delta\chi}}{s}=\frac{n_{\Delta\chi}}{\frac{2\pi^2}{45}g_*(T)T^3}= & \frac{45}{2\pi^2g_*(T)}\frac{1}{3}\left(\frac{3n_{\Delta\chi}}{T^3}\right), \nonumber 
\end{align}
we can write the chemical potential as
\begin{align}
    \mu/T &= \frac{2\pi^2g_*(T)}{15} Y_{\Delta\chi}=\mathcal{F}(T,Y_{\Delta\chi}).
\end{align}
Using the bias potential derived above, the annihilation temperature of DWs can be estimated from
\begin{align}
    \Delta V_T &= C_{\rm ann} \frac{\mathcal{A}\sigma}{t_{ann}}
\end{align}
This can be simplified in terms of the Hubble parameter in radiation dominated Universe as
\begin{align}
     & \implies H_{\rm ann} \leq \frac{\Delta V_T}{2 C_{\rm ann}\mathcal{A}\sigma}\nonumber\\
    & \implies 1.66 \sqrt{g_*(T)}\frac{1}{M_{\rm pl}}T^2\leq \frac{m_0 y v_\phi}{v_\phi^3} \frac{\mathcal{G}(\mathcal{F}(T,Y_{\Delta\chi}))  }{0.94\pi^2 \sqrt{\lambda} C_{\rm ann}\mathcal{A}}T^2 \nonumber\\
    & \implies 1.66 \sqrt{g_*(T)}\frac{1}{M_{\rm pl}}\leq C_1 \frac{\mathcal{G}(\mathcal{F}(T,Y_{\Delta\chi}))}{0.94\pi^2 \sqrt{\lambda}C_{\rm ann}\mathcal{A}}
\end{align}
where, $C_1=\frac{m_0 y v_\phi}{v_\phi^3}=\frac{m_0 y}{v_\phi^2}$ is a parameter related to the model parameters. The bias term can dominate the tension force of the DWs once a minimum amount of asymmetry ($Y_{\Delta\chi}^{min}$) is generated or injected into the bath. This requirement can be expressed as
\begin{align}
   1.66 \sqrt{g_*(T)}\frac{1}{M_{\rm pl}}= C_1 \frac{  \mathcal{G}(\mathcal{F}(T,Y_{\Delta\chi}^{\rm min}))}{0.94\pi^2 \sqrt{\lambda}C_{\rm ann}\mathcal{A}}.
\end{align}
For $Y_{\Delta\chi}^{\rm min}$=0.16, $C_1=10^{-22}$. So, DWs will annihilate if $Y_{\Delta\chi}^{\rm min}$ amount of asymmetry is injected.

\bibliographystyle{JHEP}
\bibliography{arxiv_v1.bbl}

\providecommand{\href}[2]{#2}\begingroup\raggedright\begin{thebibliography}{10}

\bibitem{Zeldovich:1974uw}
Y.B.~Zeldovich, I.Y.~Kobzarev and L.B.~Okun, \emph{{Cosmological Consequences of the Spontaneous Breakdown of Discrete Symmetry}}, {\emph{Zh. Eksp. Teor. Fiz.} {\bfseries 67} (1974) 3}.

\bibitem{Vilenkin:1981zs}
A.~Vilenkin, \emph{{Gravitational Field of Vacuum Domain Walls and Strings}}, \href{https://doi.org/10.1103/PhysRevD.23.852}{\emph{Phys. Rev. D} {\bfseries 23} (1981) 852}.

\bibitem{Sikivie:1982qv}
P.~Sikivie, \emph{{Of Axions, Domain Walls and the Early Universe}}, \href{https://doi.org/10.1103/PhysRevLett.48.1156}{\emph{Phys. Rev. Lett.} {\bfseries 48} (1982) 1156}.

\bibitem{Gelmini:1988sf}
G.B.~Gelmini, M.~Gleiser and E.W.~Kolb, \emph{{Cosmology of Biased Discrete Symmetry Breaking}}, \href{https://doi.org/10.1103/PhysRevD.39.1558}{\emph{Phys. Rev. D} {\bfseries 39} (1989) 1558}.

\bibitem{Larsson:1996sp}
S.E.~Larsson, S.~Sarkar and P.L.~White, \emph{{Evading the cosmological domain wall problem}}, \href{https://doi.org/10.1103/PhysRevD.55.5129}{\emph{Phys. Rev. D} {\bfseries 55} (1997) 5129} [\href{https://arxiv.org/abs/hep-ph/9608319}{{\ttfamily hep-ph/9608319}}].

\bibitem{Rai:1992xw}
B.~Rai and G.~Senjanovic, \emph{{Gravity and domain wall problem}}, \href{https://doi.org/10.1103/PhysRevD.49.2729}{\emph{Phys. Rev. D} {\bfseries 49} (1994) 2729} [\href{https://arxiv.org/abs/hep-ph/9301240}{{\ttfamily hep-ph/9301240}}].

\bibitem{Lew:1993yt}
H.~Lew and A.~Riotto, \emph{{Baryogenesis, domain walls and the role of gravity}}, \href{https://doi.org/10.1016/0370-2693(93)90930-G}{\emph{Phys. Lett. B} {\bfseries 309} (1993) 258} [\href{https://arxiv.org/abs/hep-ph/9304203}{{\ttfamily hep-ph/9304203}}].

\bibitem{Zhang:2023nrs}
Z.~Zhang, C.~Cai, Y.-H.~Su, S.~Wang, Z.-H.~Yu and H.-H.~Zhang, \emph{{Nano-Hertz gravitational waves from collapsing domain walls associated with freeze-in dark matter in light of pulsar timing array observations}}, \href{https://doi.org/10.1103/PhysRevD.108.095037}{\emph{Phys. Rev. D} {\bfseries 108} (2023) 095037} [\href{https://arxiv.org/abs/2307.11495}{{\ttfamily 2307.11495}}].

\bibitem{Zeng:2025zjp}
Q.-Q.~Zeng, X.~He, Z.-H.~Yu and J.~Zheng, \emph{{Collapsing domain walls with Z2-violating coupling to thermalized fermions and their impact on gravitational wave detections}}, \href{https://doi.org/10.1103/cdsj-bmvx}{\emph{Phys. Rev. D} {\bfseries 111} (2025) 115017} [\href{https://arxiv.org/abs/2501.10059}{{\ttfamily 2501.10059}}].

\bibitem{Borah:2025bfa}
D.~Borah and I.~Saha, \emph{{Gravitational waves from seesaw assisted collapsing domain walls}},  \href{https://arxiv.org/abs/2512.22339}{{\ttfamily 2512.22339}}.

\bibitem{Borah:2026kfo}
D.~Borah, P.K.~Paul and N.~Sahu, \emph{{Can Dirac neutrinos destabilize $\mathcal{Z}_2$ domain wall network?}},  \href{https://arxiv.org/abs/2602.07380}{{\ttfamily 2602.07380}}.

\bibitem{Ma:2025bjf}
E.~Ma, P.K.~Paul and N.~Sahu, \emph{{Lepton parity dark matter and naturally unstable domain walls}}, \href{https://doi.org/10.1103/tj6t-dyqn}{\emph{Phys. Rev. D} {\bfseries 112} (2025) 095020} [\href{https://arxiv.org/abs/2508.02642}{{\ttfamily 2508.02642}}].

\bibitem{Planck:2018vyg}
{\scshape Planck} collaboration, \emph{{Planck 2018 results. VI. Cosmological parameters}}, \href{https://doi.org/10.1051/0004-6361/201833910}{\emph{Astron. Astrophys.} {\bfseries 641} (2020) A6} [\href{https://arxiv.org/abs/1807.06209}{{\ttfamily 1807.06209}}].

\bibitem{Kibble:1976sj}
T.W.B.~Kibble, \emph{{Topology of Cosmic Domains and Strings}}, \href{https://doi.org/10.1088/0305-4470/9/8/029}{\emph{J. Phys. A} {\bfseries 9} (1976) 1387}.

\bibitem{Saikawa:2017hiv}
K.~Saikawa, \emph{{A review of gravitational waves from cosmic domain walls}}, \href{https://doi.org/10.3390/universe3020040}{\emph{Universe} {\bfseries 3} (2017) 40} [\href{https://arxiv.org/abs/1703.02576}{{\ttfamily 1703.02576}}].

\bibitem{Roshan:2024qnv}
R.~Roshan and G.~White, \emph{{Using gravitational waves to see the first second of the Universe}},  \href{https://arxiv.org/abs/2401.04388}{{\ttfamily 2401.04388}}.

\bibitem{Coleman:1973jx}
S.R.~Coleman and E.J.~Weinberg, \emph{{Radiative Corrections as the Origin of Spontaneous Symmetry Breaking}}, \href{https://doi.org/10.1103/PhysRevD.7.1888}{\emph{Phys. Rev. D} {\bfseries 7} (1973) 1888}.

\bibitem{Hiramatsu:2013qaa}
T.~Hiramatsu, M.~Kawasaki and K.~Saikawa, \emph{{On the estimation of gravitational wave spectrum from cosmic domain walls}}, \href{https://doi.org/10.1088/1475-7516/2014/02/031}{\emph{JCAP} {\bfseries 02} (2014) 031} [\href{https://arxiv.org/abs/1309.5001}{{\ttfamily 1309.5001}}].

\bibitem{Hiramatsu:2012sc}
T.~Hiramatsu, M.~Kawasaki, K.~Saikawa and T.~Sekiguchi, \emph{{Axion cosmology with long-lived domain walls}}, \href{https://doi.org/10.1088/1475-7516/2013/01/001}{\emph{JCAP} {\bfseries 01} (2013) 001} [\href{https://arxiv.org/abs/1207.3166}{{\ttfamily 1207.3166}}].

\bibitem{Kadota:2015dza}
K.~Kadota, M.~Kawasaki and K.~Saikawa, \emph{{Gravitational waves from domain walls in the next-to-minimal supersymmetric standard model}}, \href{https://doi.org/10.1088/1475-7516/2015/10/041}{\emph{JCAP} {\bfseries 10} (2015) 041} [\href{https://arxiv.org/abs/1503.06998}{{\ttfamily 1503.06998}}].

\bibitem{Chen:2020wvu}
N.~Chen, T.~Li and Y.~Wu, \emph{{The gravitational waves from the collapsing domain walls in the complex singlet model}}, \href{https://doi.org/10.1007/JHEP08(2020)117}{\emph{JHEP} {\bfseries 08} (2020) 117} [\href{https://arxiv.org/abs/2004.10148}{{\ttfamily 2004.10148}}].

\bibitem{Bhattacharya:2023kws}
S.~Bhattacharya, N.~Mondal, R.~Roshan and D.~Vatsyayan, \emph{{Leptogenesis, dark matter and gravitational waves from discrete symmetry breaking}}, \href{https://doi.org/10.1088/1475-7516/2024/06/029}{\emph{JCAP} {\bfseries 06} (2024) 029} [\href{https://arxiv.org/abs/2312.15053}{{\ttfamily 2312.15053}}].

\bibitem{Paul:2024iie}
P.K.~Paul, N.~Sahu and P.~Shukla, \emph{{Thermal leptogenesis, dark matter, and gravitational waves from an extended canonical seesaw scenario}}, \href{https://doi.org/10.1103/w8gl-wbjd}{\emph{Phys. Rev. D} {\bfseries 112} (2025) 015032} [\href{https://arxiv.org/abs/2409.08828}{{\ttfamily 2409.08828}}].

\bibitem{Caprini:2018mtu}
C.~Caprini and D.G.~Figueroa, \emph{{Cosmological Backgrounds of Gravitational Waves}}, \href{https://doi.org/10.1088/1361-6382/aac608}{\emph{Class. Quant. Grav.} {\bfseries 35} (2018) 163001} [\href{https://arxiv.org/abs/1801.04268}{{\ttfamily 1801.04268}}].

\bibitem{2017arXiv170200786A}
{\scshape LISA} collaboration, \emph{{Laser Interferometer Space Antenna}}, {\emph{arXiv e-prints} (2017) arXiv:1702.00786} [\href{https://arxiv.org/abs/1702.00786}{{\ttfamily 1702.00786}}].

\bibitem{Punturo_2010}
{\scshape ET Collaboration} collaboration, \emph{The einstein telescope: a third-generation gravitational wave observatory}, \href{https://doi.org/10.1088/0264-9381/27/19/194002}{\emph{Classical and Quantum Gravity} {\bfseries 27} (2010) 194002}.

\bibitem{Garcia-Bellido:2021zgu}
J.~Garcia-Bellido, H.~Murayama and G.~White, \emph{{Exploring the Early Universe with Gaia and THEIA}},  \href{https://arxiv.org/abs/2104.04778}{{\ttfamily 2104.04778}}.

\bibitem{Sesana:2019vho}
A.~Sesana et~al., \emph{{Unveiling the gravitational universe at $\mu$-Hz frequencies}}, \href{https://doi.org/10.1007/s10686-021-09709-9}{\emph{Exper. Astron.} {\bfseries 51} (2021) 1333} [\href{https://arxiv.org/abs/1908.11391}{{\ttfamily 1908.11391}}].

\bibitem{Weltman:2018zrl}
A.~Weltman et~al., \emph{{Fundamental physics with the Square Kilometre Array}}, \href{https://doi.org/10.1017/pasa.2019.42}{\emph{Publ. Astron. Soc. Austral.} {\bfseries 37} (2020) e002} [\href{https://arxiv.org/abs/1810.02680}{{\ttfamily 1810.02680}}].

\bibitem{Adelberger:2005bt}
E.G.~Adelberger, N.A.~Collins and C.D.~Hoyle, \emph{{Analytic expressions for gravitational inner multipole moments of elementary solids and for the force between two rectangular solids}}, \href{https://doi.org/10.1088/0264-9381/23/1/007}{\emph{Class. Quant. Grav.} {\bfseries 23} (2006) 125} [\href{https://arxiv.org/abs/gr-qc/0512055}{{\ttfamily gr-qc/0512055}}].

\bibitem{NANOGrav:2023gor}
{\scshape NANOGrav} collaboration, \emph{{The NANOGrav 15 yr Data Set: Evidence for a Gravitational-wave Background}}, \href{https://doi.org/10.3847/2041-8213/acdac6}{\emph{Astrophys. J. Lett.} {\bfseries 951} (2023) L8} [\href{https://arxiv.org/abs/2306.16213}{{\ttfamily 2306.16213}}].

\bibitem{CMB-HD:2022bsz}
{\scshape CMB-HD} collaboration, \emph{{Snowmass2021 CMB-HD White Paper}},  \href{https://arxiv.org/abs/2203.05728}{{\ttfamily 2203.05728}}.

\bibitem{Barman:2022yos}
B.~Barman, D.~Borah, A.~Dasgupta and A.~Ghoshal, \emph{{Probing High Scale Dirac Leptogenesis via Gravitational Waves from Domain Walls}},  \href{https://arxiv.org/abs/2205.03422}{{\ttfamily 2205.03422}}.

\bibitem{Barman:2023fad}
B.~Barman, D.~Borah, S.~Jyoti~Das and I.~Saha, \emph{{Scale of Dirac leptogenesis and left-right symmetry in the light of recent PTA results}}, \href{https://doi.org/10.1088/1475-7516/2023/10/053}{\emph{JCAP} {\bfseries 10} (2023) 053} [\href{https://arxiv.org/abs/2307.00656}{{\ttfamily 2307.00656}}].

\bibitem{King:2023cgv}
S.F.~King, D.~Marfatia and M.H.~Rahat, \emph{{Towards distinguishing Dirac from Majorana neutrino mass with gravitational waves}},  \href{https://arxiv.org/abs/2306.05389}{{\ttfamily 2306.05389}}.

\bibitem{Escudero:2022okz}
M.~Escudero, A.~Ibarra and V.~Maura, \emph{{Primordial lepton asymmetries in the precision cosmology era: Current status and future sensitivities from BBN and the CMB}}, \href{https://doi.org/10.1103/PhysRevD.107.035024}{\emph{Phys. Rev. D} {\bfseries 107} (2023) 035024} [\href{https://arxiv.org/abs/2208.03201}{{\ttfamily 2208.03201}}].

\bibitem{Borah:2024xoa}
D.~Borah, N.~Das and I.~Saha, \emph{{Large neutrino asymmetry from forbidden decay of dark matter}}, \href{https://doi.org/10.1103/x2z2-whhf}{\emph{Phys. Rev. D} {\bfseries 112} (2025) 115044} [\href{https://arxiv.org/abs/2410.00096}{{\ttfamily 2410.00096}}].

\bibitem{Matsumoto:2022tlr}
A.~Matsumoto et~al., \emph{{EMPRESS. VIII. A New Determination of Primordial He Abundance with Extremely Metal-Poor Galaxies: A Suggestion of the Lepton Asymmetry and Implications for the Hubble Tension}},  \href{https://arxiv.org/abs/2203.09617}{{\ttfamily 2203.09617}}.

\bibitem{Burns:2022hkq}
A.-K.~Burns, T.M.P.~Tait and M.~Valli, \emph{{Indications for a Nonzero Lepton Asymmetry in the Early Universe}},  \href{https://arxiv.org/abs/2206.00693}{{\ttfamily 2206.00693}}.

\bibitem{Borah:2026xxx}
D.~Borah, P.K.~Paul and N.~Sahu, \emph{{Cogenesis of visible and dark matter in type-I Dirac seesaw}},  \href{https://arxiv.org/abs/2603.24693}{{\ttfamily 2603.24693}}.

\bibitem{Borah:2025dka}
D.~Borah and A.~Dasgupta, \emph{{Electromagnetic Dirac Cogenesis}},  \href{https://arxiv.org/abs/2507.11607}{{\ttfamily 2507.11607}}.

\bibitem{Bandyopadhyay:2025hoc}
D.~Bandyopadhyay, D.~Borah and A.~Dasgupta, \emph{{ALPy Cogenesis}},  \href{https://arxiv.org/abs/2506.18970}{{\ttfamily 2506.18970}}.

\bibitem{Borah:2022uos}
D.~Borah and A.~Dasgupta, \emph{{Large neutrino asymmetry from TeV scale leptogenesis}}, \href{https://doi.org/10.1103/PhysRevD.108.035015}{\emph{Phys. Rev. D} {\bfseries 108} (2023) 035015} [\href{https://arxiv.org/abs/2206.14722}{{\ttfamily 2206.14722}}].

\bibitem{Affleck:1984fy}
I.~Affleck and M.~Dine, \emph{{A New Mechanism for Baryogenesis}}, \href{https://doi.org/10.1016/0550-3213(85)90021-5}{\emph{Nucl. Phys. B} {\bfseries 249} (1985) 361}.

\end{thebibliography}\endgroup

\end{document}